\documentclass[12pt]{article}
\usepackage{graphicx}
\usepackage{amsmath}
\usepackage{amssymb}
\usepackage{ulem}
\usepackage{enumerate}
\usepackage{cite}
\usepackage{latexsym}
\begin{document}
\begin{center}
{\bf {\large{Continuous and Discrete Symmetries in a  4D Field-Theoretic System:  Symmetry Operators and Their Algebraic Structures}}}

\vskip 3.4cm

{\sf  R. P. Malik$^{(a,b)}$}\\
$^{(a)}$ {\it Physics Department, Institute of Science,}\\
{\it Banaras Hindu University, Varanasi - 221 005, India}\\

\vskip 0.1cm

$^{(b)}$ {\it DST Centre for Interdisciplinary Mathematical Sciences,}\\
{\it Institute of Science, Banaras Hindu University, Varanasi - 221 005, India}\\
{\small {\sf {e-mails: rpmalik1995@gmail.com; malik@bhu.ac.in}}}
\end{center}

\vskip 3.0 cm

\noindent
{\bf Abstract:}
Within the framework of Becchi-Rouet-Stora-Tyutin (BRST) formalism, we show the existence of (i) a couple of off-shell nilpotent (i.e. fermionic)
BRST and co-BRST symmetry transformations, and (ii) a full set of  non-nilpotent (i.e. bosonic) symmetry transformations for an
appropriate  Lagrangian density that describes 
the {\it combined} system of the free Abelian 3-form and 1-form gauge theories in the physical four (3 + 1)-dimensions of the flat Minkowskian  spacetime.
This combined BRST-quantized field-theoretic system is essential for the existence of the 
off-shell nilpotent co-BRST and non-nilpotent bosonic symmetry transformations in the theory.
We concentrate on the full algebraic structures of the above continuous symmetry transformation operators {\it along}
with a couple of very useful discrete duality symmetry transformation operators existing in our
four (3 + 1)-dimensional (4D) field-theoretic model. We establish the relevance of the algebraic structures, respected by  the {\it above} 
discrete and continuous symmetry operators, to the algebraic structures that are obeyed by the de Rham cohomological  operators of 
differential geometry. 
One of the highlights of our present endeavor is the observation that there are no ``exotic'' fields with the negative kinetic terms
in our present 4D {\it field-theoretic} example for Hodge theory. 



\newpage

\section {Introduction}

The research activities, related to the ideas behind (super)string theories (see, e.g. [1-3] and references therein), are the forefront areas of genuine 
interest in 
the modern-day theoretical high energy physics (THEP). One of the key consequences of the quantum excitations of (super)strings has been the observation
that the higher $p$-form ($p = 2, 3, ...$) basic fields appear in {\it these} excitations which, very naturally, push the (super)string theories to go beyond 
the realm of the standard model of elementary particle physics that is based on the non-Abelian 1-form (i.e. $p = 1$) 
{\it interacting} gauge theory. Hence, there
has been interest in the study of the gauge theories that are based on the higher $p$-form ($p = 2, 3, ...$) {\it basic} gauge fields which have very rich mathematical
and physical structures. Our present endeavor is a modest step in that direction where we study the physical four (3 + 1)-dimensional (4D) 
{\it combined} field-theoretic
system of the free Abelian 3-form and 1-form gauge theories within the framework of Becchi-Rouet-Stora-Tyutin (BRST) formalism [4-7].

Our present investigation is essential on the following counts. First of all, we have been able to establish that the 4D {\it massless} and 
the St{\" u}ckelberg-modified {\it massive} Abelian 2-form 
BRST-quantized gauge theories are the field-theoretic examples for Hodge theory [8,9]. In our present endeavor, we propose a {\it new} 4D 
BRST-quantized field-theoretic model which is {\it also} an example for Hodge theory. Second, in our earlier work\footnote{We have established
 that the $2p$-dimensional  St{\" u}ckelberg-modified {\it massive} Abelian $p$-form ($ p = 1, 2, 3$) BRST-quantized 
gauge theories are the field-theoretic examples for Hodge theory with a tower of $p$-number of ``exotic'' fields. However, it has {\it not} been 
clear as to which ``exotic'' field (from the above $p$-number of fields) is
 the most fundamental one. In view of our {\it earlier} work [9], our {\it present} work makes it 
clear that the pseudo-scalar field is the most fundamental ``exotic'' field in the {\it physical} four dimensions of spacetime. On
the contrary, the axial-vector field of [9] is {\it not} a genuine ``exotic'' field (cf. section five, too).} on the 4D model [9], we
have been able to show the existence of an axial-vector and a pseudo-scalar 
``exotic'' fields with the negative kinetic terms\footnote{Such kinds of fields with the negative kinetic terms ({\it with} and {\it without} rest masses)
have  also been considered as a set of possible candidates for dark matter/dark energy (see, e.g. [10,11] for details).} 
which are a set of possible candidates
for the phantom fields of the cosmological models (see, e.g. [12-14] and references therein). In our present endeavor, we demonstrate that there is
{\it no} existence of any kinds of ``exotic'' fields with the negative kinetic terms. 
Third, we show that the BRST-quantized Lagrangian densities of the {\it combined} Abelian 3-form and 1-form gauge theories remain invariant, 
separately and independently, under the BRST transformations.  However, for the invariance of the 
co-BRST transformations, we need
{\it both} of them {\it together} in one field-theoretic system. Finally, we focus on the algebraic structures that are satisfied by the 
discrete and continuous symmetry
operators and establish their resemblance with the Hodge algebra that is satisfied by the de Rham cohomological operators 
of differential geometry (see, e.g. [15,16]).

Against the backdrop of the above paragraph, it is crystal clear that our present 4D field-theoretic system of Hodge theory rules out the
(axial-)vector fields to be  a set of possible candidates for the phantom fields of the cosmological models of the Universe. Thus, as far
as our earlier work [9] on the 4D St{\" u}ckelberg-modified massive Abelian 2-form theory is concerned, it is now established that 
out of the axial-vector field and the pseudo-scalar field, the pseudo-scalar (PS) field is the most fundamental object that corresponds to 
a  possible candidate for the phantom field of the cosmological models. This is backed by our observations that this ``exotic''
PS field (with the negative kinetic term) appears in (i) the 2D  St{\" u}ckelberg-modified  massive Abelian 1-form (i.e. Proca)  theory 
(see, e.g. [17] and references therein),
and (ii) the 3D field-theoretic system of the combination of the free Abelian 2-form and 1-form gauge theories
(see, e.g. [18] and references therein). Both {\it these}
2D and 3D field-theoretic systems {\it also} provide  a set of examples for Hodge theory.

The theoretical contents of our present investigation are organized as follows. In the next section, we define the proper gauge-fixed 
preliminary {\it classical} Lagrangian density
for our {\it combined} system of the free 4D Abelian 3-form and 1-form gauge theories. Our section three is devoted to the elevation of the {\it most} general
classical gauge-fixed 
Lagrangian density to its {\it quantum} counterpart (i.e. the (co-)BRST invariant Lagrangian density) that incorporates the Faddeev-Popov (FP) ghost terms
where we also pinpoint the existence of a couple of discrete duality symmetry transformations and their usefulness in the algebraic structures 
that are obeyed by the symmetry operators of our theory. 
In our section four, we deal with a bosonic symmetry operator that is derived from the anticommutator of the nilpotent (co-) BRST symmetry transformation operators
where we {\it also} discuss the algebraic structures 
that are obeyed by the discrete as well as the continuous symmetry transformation operators of our theory. Finally,
in our section five, we summarize our key results and point out the future perspective and scope of our present investigation.\\

\vskip 0.7cm

\section{Preliminaries: Gauge-Fixed Lagrangian Densities}

In the {\it physical} four (3 + 1)-dimensional (4D) spacetime, we have the following standard form of the starting Lagrangian density 
(${\cal L}_{(0)} $) for the {\it combined} field-theoretic system of the {\it free} Abelian 3-form and 1-form gauge theories\footnote{
We adopt the convention of the left derivative w.r.t. all the {\it fermionic} fields
of our theory. We take the 4D flat Minkowskian metric tensor $\eta_{\mu\nu}$ as: $\eta_{\mu\nu}$ = 
diag $(+ 1, -1, -1, -1)$ so that the dot product between two {\it non-null} 4D vectors $P_\mu$ and $Q_\mu$ is defined as: 
$P\cdot Q = \eta_{\mu\nu} P^\mu\,Q^\nu \equiv P_0\, Q_0 - P_i\,Q_i$ where the Greek indices 
$ \mu, \nu, \sigma ... = 0, 1, 2,  3$ stand for the time and space directions and Latin indices $i, j, k... = 1, 2, 3$ correspond to 
the 3D space directions {\it only}. 
The 4D Levi-Civita tensor $\varepsilon_{\mu\nu\sigma\rho}$ is chosen such that  $\varepsilon_{0123} 
= +1 = - \varepsilon^{0123}$ and they satisfy the standard relationships: $\varepsilon_{\mu\nu\eta\kappa} \varepsilon^{\mu\nu\eta\kappa}= - \,4!$,
$\varepsilon_{\mu\nu\eta\kappa} \varepsilon^{\mu\nu\eta\rho}= - \,3! \,\delta^\rho_\kappa, \; \varepsilon_{\mu\nu\eta\kappa} \varepsilon^{\mu\nu\sigma\rho}= - \,2!\,
\big(\delta^\sigma_\eta \delta^\rho_\kappa - \delta^\sigma_\kappa \delta^\rho_\eta\big)$, etc. 
We also adopt the convention: $(\delta A_{\mu\nu\sigma}/\delta A_{\alpha\beta\gamma}) = \frac{1}{3!}\,\big[ \delta^\alpha_\mu (\delta^\beta_\nu \,
\delta^\gamma_\sigma - \delta^\beta_\sigma \,\delta^\gamma_\nu) +  \delta^\alpha_\nu (\delta^\beta_\sigma \,
\delta^\gamma_\mu - \delta^\beta_\mu \,\delta^\gamma_\sigma) + \delta^\alpha_\sigma (\delta^\beta_\mu \,
\delta^\gamma_\nu - \delta^\beta_\nu \,\delta^\gamma_\mu)\big]$, etc., for the tensorial dififferentiation/variation
for various computational purposes.} (see. e.g. [19] for details):
\begin{eqnarray}\label{1}
{\cal L}_{(0)} &=& \dfrac{1}{48}\, H^{\mu\nu\sigma\rho}\, H_{\mu\nu\sigma\rho}  - \dfrac{1}{4}\, F^{\mu\nu}\, F_{\mu\nu} = -\,\dfrac{1}{2}\,(H_{0123})^2 
- \dfrac{1}{4}\, (F_{\mu\nu})^2 
\nonumber\\ 
&\equiv&   
-\, \dfrac{1}{2}\,\Big(- \dfrac{1}{3!}\varepsilon^{\mu\nu\sigma\rho} \,\partial_\mu A_{\nu\sigma\rho} \Big)^2
+ \dfrac{1}{4}\, \Big( \varepsilon^{\mu\nu\sigma\rho} \,\partial_\sigma A_\rho \Big)^2.
\end{eqnarray}
Here the field-strength tensor $H_{\mu \nu \sigma \rho} = \partial_\mu\, A_{\nu\sigma\rho} - \partial_\nu\,
 A_{\sigma\rho \mu } + \partial_\sigma\, A_{\rho \mu \nu }  - \partial_\rho \,A_{\mu \nu \sigma }$
is derived from the 4-form $H^{(4)} = d\, A^{(3)}$ where $A^{(3)} = \frac{1}{3!}\,A_{\mu\nu\sigma}\,(d\,x^\mu \wedge d\,x^\nu \wedge d\, x^\sigma)$
defines the {\it totally} antisymmetric tensor 
(i.e. Abelian 3-form) gauge field $A_{\mu\nu\sigma}$. In the above,  the operator $d $ (with $d^2 = 0)$
is the exterior derivative of differential geometry (see, e.g. [15,16] for details) and the explicit form of $H^{(4)}$ is: $H^{(4)} = d\, A^{(3)} 
= \frac{1}{4!}\,\, H_{\mu \nu \sigma\rho}\, ( d\,x^\mu \wedge d\,x^\nu \wedge d\, x^\sigma \wedge d\, x^\rho)$. In exactly similar fashion, the 
Abelian 2-form: $F^{(2)} = d\, A^{(1)} \equiv \frac{1}{2!}\, F_{\mu\nu} \big(dx^\mu \wedge dx^\nu \big)$ defines the field-strength tensor
$F_{\mu\nu}  = \partial_\mu A_\nu - \partial_\nu A_\mu $ for the Abelian 1-form (i.e. $A^{(1)} = A_\mu\, dx^\mu $) gauge field $A_\mu$. It is the {\it special}
feature of our 4D theory that (i) the kinetic terms for the Abelian 3-form and 1-form gauge fields 
are expressed in terms of the 4D Levi-Civita tensor, (ii)
the field-strength tensor of the Abelian 3-form 
gauge field has only a {\it single} existing independent component because we observe that the general form  of the kinetic term
for {\it this} field is: $\frac{1}{48}\, H^{\mu\nu\sigma\rho}\, H_{\mu\nu\sigma\rho} = \frac{1}{2}\, H^{0123}\, H_{0123} \equiv -\,\frac{1}{2}\, (H_{0123})^2 $, 
and (iii) the covariant forms of the existing components of the field-strength tensor for the Abelian 3-from gauge field ($A_{\mu\nu\sigma}$) are: 
$H^{0123} = +\,\frac{1}{3!}\, \varepsilon_{\mu\nu\sigma\rho}\, \partial^\mu A^{\nu\sigma\rho}$ and 
$H_{0123} = -\,  \frac{1}{3!}\, \varepsilon^{\mu\nu\sigma\rho}\, \partial_\mu A_{\nu\sigma\rho}$.

The 4D theory, described by the Lagrangian density (1), is endowed with a set of first-class constraints in the terminology of Dirac's prescription for
the classification scheme of constraints (see, e.g. [20,21] for details). These constraints generate the infinitesimal, local and continuous gauge symmetry
transformations: $\delta_g A_{\mu\nu\sigma} =  \partial_\mu \Lambda_{\nu\sigma} + \partial_\nu \Lambda_{\sigma\mu} 
+ \partial_\sigma \Lambda_{\mu\nu}, \; \delta_g A_\mu = \partial_\mu\Lambda $ under which the kientic terms 
for {\it both} the gauge fields remain invariant (and, hence, the
Lagrangian density (1), too). Here the antisymmetric (i.e. $\Lambda_{\mu\nu} = -\, \Lambda_{\nu\mu} $) tensor 
$\Lambda_{\mu\nu} $ and Lorentz scalar $\Lambda$ 
are the infinitesimal local gauge symmetry transformation parameters [cf. Eq. (4) below]. 
To quantize this theory, we need to add the {\it proper} gauge-fixing terms. At a very {\it preliminary} level, we have the 
following forms (i.e. ${\cal L}_{(1)} $) 
of the gauge-fixed Lagrangian density (which are the {\it equivalent} generalizations of (1)), namely;
\begin{eqnarray}\label{2}
{\cal L}_{(1)}  &=& -\, 
\dfrac{1}{2}\,\Big(\dfrac{1}{3!}\varepsilon^{\mu\nu\sigma\rho} \,\partial_\mu A_{\nu\sigma\rho} \Big)^2
+ \dfrac{1}{4} \big(\partial^\nu A_{\nu\mu\sigma} \big)^2  
+ \dfrac{1}{4} \,\Big(\, \varepsilon^{\mu\nu\sigma\rho} \,\partial_\sigma A_\rho \Big)^2
- \dfrac{1}{2} \big(\partial \cdot A \big)^2 \nonumber\\
&\equiv& \dfrac{1}{48}\, H^{\mu\nu\sigma\rho}\, H_{\mu\nu\sigma\rho}  + \dfrac{1}{4} \big(\partial^\nu A_{\nu\mu\sigma} \big)^2  
 - \dfrac{1}{4}\, F^{\mu\nu}\, F_{\mu\nu} - \dfrac{1}{2} \big(\partial \cdot A \big)^2.  
\end{eqnarray}
A few noteworthy points, at this juncture, are as follows. First of all, we note that the top entry in (2) is valid only when our theory is defined 
on the 4D flat Minkowskian spacetime manifold. On the other hand, the bottom entry in 
equation (2) is valid in any arbitrary D-dimension of spacetime (including the 4D spacetime). Second,
the gauge-fixing terms in (2) owe their origin to the co-exterior derivative $\delta = -\, *\, d\, *$ (with $\delta^2 = 0 $) of differential geometry [15,16]
on the 4D spacetime manifold because we observe that: $\delta \, A^{(1)} = +\, (\partial \cdot A)$ and
$\delta \, A^{(3)} = -\, \frac{1}{2!}\, (\partial^\nu A_{\nu\sigma\mu})\, (dx^\sigma \wedge dx^\mu)$.
Here the symbol $*$ stands for the Hodge duality operator on the flat 4D spacetime that has been chosen for our theoretical discussions.
Third, it is straightforward  to check that we obtain the Euler-Lagrange (EL) equations of motion (EoM): $\Box A_{\mu\nu\sigma} = 0, \; \Box A_\mu = 0 $
(for the {\it massless} gauge fields $A_{\mu\nu\sigma}$ and $A_\mu$)
from the bottom entry of the
above gauge-fixed Lagrangian density\footnote{From the top entry of the
gauge-fixed Lagrangian density (2), it is clear that
we shall obtain the EL-EoM for the $A_\mu$ field as: $\frac{1}{2}\, \varepsilon^{\mu\nu\sigma\rho} \,\varepsilon_{\sigma\rho\eta\kappa}\, \partial_\mu \,
\partial^\eta A^\kappa - \partial^\nu (\partial \cdot A) = 0 $ which, ultimatley, leads to $\Box A_\mu = 0 $ provided we use the 
standard relationship: $\varepsilon^{\mu\nu\eta\kappa} \varepsilon_{\mu\nu\sigma\rho}= - \,2!\,
\big(\delta_\sigma^\eta \delta_\rho^\kappa - \delta_\sigma^\kappa \delta_\rho^\eta\big) $. In exactly similar fashion, we obseve that
the EL-EoM for the Abelian 3-form gauge field is:$ -\, \frac{1}{3!}\,\varepsilon^{\mu\nu\sigma\rho}\, \partial_\mu (\varepsilon^{\alpha\beta\gamma\delta}
\, \partial_\alpha A_{\beta\gamma\delta} ) + \partial^\nu (\partial_\eta A^{\eta\sigma\rho}) + \partial^\sigma (\partial_\eta A^{\eta\rho\nu})
+ \partial^\rho (\partial_\eta A^{\eta\nu\sigma}) = 0$. Using the relationship: $-\, 3!\, H_{0123} = \varepsilon^{\alpha\beta\gamma\delta}\, \partial_\alpha A_{\beta\gamma\delta} $, we can recast {\it this} EL-EoM as:  $ \varepsilon^{\mu\nu\sigma\rho}\, \partial_\mu (H_{0123}) + \partial^\nu (\partial_\eta A^{\eta\sigma\rho}) + \partial^\sigma (\partial_\eta A^{\eta\rho\nu}) + \partial^\rho (\partial_\eta A^{\eta\nu\sigma}) = 0$ which leads to 
$\Box A_{\mu\nu\sigma} = 0 $ [where we have $A_{\mu\nu\sigma} = (A_{012},\, A_{123},\, A_{301},\, A_{230}) $  for our 4D field-theoretic model]. }. 
Finally, we note that under the following discrete duality\footnote{The mathematical
basis for (i) the symmetry transformations (3), and (ii) the numerical factors appearing therein, can be explained (modulo a factor of $\pm$ signs) by taking
into account the Hodge duality $*$ operation on our chosen 4D 
flat spacetime manifold because we observe that: $*\, A^{(1)} = *\, (A_\mu \, dx^\mu)
= \frac{1}{3!}\, \varepsilon_{\mu\nu\sigma\rho}\, A^\mu\, (dx^\nu \wedge dx^\sigma \wedge dx^\rho) \sim \frac{1}{3!}\,
A_{\nu\sigma\rho}\, (dx^\nu \wedge dx^\sigma \wedge dx^\rho)$
and $*\, A^{(3)} = *\, \big[\frac{1}{3!}\,A_{\nu\sigma\rho}\, (dx^\nu \wedge dx^\sigma \wedge dx^\rho) \big ] 
= \frac{1}{3!}\, \varepsilon_{\nu\sigma\rho\mu}\,A^{\nu\sigma\rho}\, (dx^\mu) \sim A_{\mu}\, dx^\mu$. This is why we call the discrete 
transformations as the {\it duality} transformations because they connect the Abelian 3-form and 1-form {\it basic} gauge fields through the 
Hodge duality $*$ operator on our 4D manifold in the sense that the {\it former} relationship implies: $A_{\mu\nu\sigma} \Rightarrow \,\pm\, \varepsilon_{\mu\nu\sigma\rho}\, A^\rho $ and {\it latter} relationship leads to: $A_\mu \Rightarrow \,\mp\, \frac{1}{3!}\, \varepsilon_{\mu\nu\sigma\rho}\, A^{\nu\sigma\rho} $ which are present in the duality transformations (3).} 
symmetry transformations\\
\begin{eqnarray}\label{3}
A_\mu \longrightarrow \,\mp\, \frac{1}{3!}\, \varepsilon_{\mu\nu\sigma\rho}\, A^{\nu\sigma\rho}, \quad 
A_{\mu\nu\sigma} \longrightarrow \,\pm\, \varepsilon_{\mu\nu\sigma\rho}\, A^\rho,
\end{eqnarray}
the kinetic term for the Abelian 3-from field interchanges with the gauge-fixing term for the Abelian 1-form field 
(i.e. $\big [-\, \frac{1}{2}\,(\frac{1}{3!}\varepsilon^{\mu\nu\sigma\rho} \,\partial_\mu A_{\nu\sigma\rho})^2
\Leftrightarrow - \frac{1}{2}\, \big(\partial \cdot A )^2 \big]$)
and the kinetic term of the
Abelian 1-from field interchanges with the gauge-fixing term of the Abelian 3-from field 
(i.e. $\big [ \frac{1}{4}\, \big(\varepsilon^{\mu\nu\sigma\rho} \,\partial_\sigma A_\rho \big)^2 
\Leftrightarrow \frac{1}{4}\, \big(\partial^\nu A_{\nu\mu\sigma} \big)^2 \big]$). In other words, the discrete duality transformstions (3) are
the {\it symmetry} transformations for the 4D gauge-fixed Lagrangian density (cf. top entry in equation (2))
for our physical 4D {\it combined} field-theoretic system of {\it two} gauge theories.

We are in the position to discuss the infinitesimal, continuous and local (dual-)gauge symmetry transformations $\delta_{(d)g}$ for the gauge-fixed
Lagrangian density ${\cal L}_{(1)} $ [cf. Eq. (2)] and obtain the mathematical restrictions  on the (dual-)gauge transformation parameters
for the symmetry invariance of the Lagrangian density (2) under {\it these} transformations. Toward this goal in mind, we note that under the 
following (dual-)gauge transformations
\begin{eqnarray}\label{4}
\delta_{dg }A_{\mu\nu\sigma} &=&   \varepsilon_{\mu\nu\sigma\rho}\, \partial^\rho \Sigma, \qquad
 \delta_{dg} A_\mu = \frac{1}{2}\;\varepsilon_{\mu\nu\sigma\rho} \, \partial^\nu \Sigma^{\sigma\rho}, \nonumber\\
\delta_g A_{\mu\nu\sigma} &=&  \partial_\mu \Lambda_{\nu\sigma} + \partial_\nu \Lambda_{\sigma\mu} 
+ \partial_\sigma \Lambda_{\mu\nu}, \qquad \delta_g A_\mu = \partial_\mu\Lambda, 
\end{eqnarray}
the Lagrangian density ${\cal L}_{(1)} $ transforms as:
\begin{eqnarray}\label{5}
\delta_{dg} {\cal L}_{(1)}  &=& -\, 
\big(\varepsilon^{\mu\nu\sigma\rho} \,\partial_\mu A_{\nu\sigma\rho} \big)\, \Box \Sigma
+ \dfrac{1}{2}\,\big(\varepsilon^{\mu\nu\sigma\rho} \,\partial_\sigma A_{\rho} \big)\, \Big [\Box \Sigma_{\mu\nu} 
- \partial_\mu \big (\partial^\eta \Sigma_{\eta\nu} \big )  + \partial_\nu \big (\partial^\eta \Sigma_{\eta\mu} \big ) \Big ], \nonumber\\
\delta_{g} {\cal L}_{(1)}  &=& \dfrac{1}{2}\, \big(\partial_\sigma A^{\sigma\mu\nu} \big)\, \Big [\Box \Lambda_{\mu\nu} 
- \partial_\mu \big (\partial^\eta \Lambda_{\eta\nu} \big ) + \partial_\nu \big (\partial^\eta \Lambda_{\eta\mu} \big ) \Big ]
- \big (\partial \cdot A \big)\, \Box \Lambda.
\end{eqnarray}
A few key and crucial points, at this stage, are in order now. First of all, we have assumed that there is parity symmetry 
invariance in the theory. As a consequence, it is clear
that the antisymmetric ($\Sigma_{\mu\nu} = -\, \Sigma_{\nu\mu} $) pseudo-tensor $\Sigma_{\mu\nu} $ 
and pseudo-scalar $\Sigma$ are the dual-gauge transformation parameters
and the transformation parameters $\Lambda_{\mu\nu}$ (with $\Lambda_{\mu\nu} = -\, \Lambda_{\nu\mu} $) 
and pure-scalar $\Lambda $ are the infinitesimal gauge transformation parameters. Second,
we note that the gauge-fixing and kinetic terms remain invariant under the (dual-)gauge symmetry transportations, respectively. Third, for the (dual-)gauge
symmetry invariance (i.e. $\delta_{(d)g} {\cal L}_{(1)} = 0 $), we have to impose exactly similar kinds of {\it outside} restrictions, namely;
\begin{eqnarray}\label{6}
 &&\Box \Sigma = 0, \qquad \Box \Sigma_{\mu\nu} 
- \partial_\mu \big (\partial^\eta \Sigma_{\eta\nu} \big )  + \partial_\nu \big (\partial^\eta \Sigma_{\eta\mu} \big ) = 0, \nonumber\\
&&\Box \Lambda = 0, \qquad \Box \Lambda_{\mu\nu} 
- \partial_\mu \big (\partial^\eta \Lambda_{\eta\nu} \big ) + \partial_\nu \big (\partial^\eta \Lambda_{\eta\mu} \big ) = 0,
\end{eqnarray}
on the (dual-)gauge transformation parameters. Finally, we shall see that there will {\it not} be any such kinds of {\it outside} restrictions
on any field when we shall discuss our present 4D field-theoretic system within the framework of BRST formalism (cf. next section).

We conclude our present section with a couple of remarks. First, the quadratic terms of the 
4D {\it preliminary} gauge-fixed Lagrangian density (2) can be linearized by invoking 
a set of Nakanishi-Lautrup type bosonic auxiliary fields ($B,\, B_1,\, B_{\mu\nu}^{(1)}, \,B_{\mu\nu}^{(2)}$). The ensuing linearized version of 
the Lagrangian density (i.e. $ {\cal L}_{(1)} \to {\cal L}_{(2)}$), namely;
\begin{eqnarray}\label{7}
{\cal L}_{(2)}  &=& \dfrac{1}{2} \, B_1^2 -\, B_1\, \Big(\dfrac{1}{3!}\varepsilon^{\mu\nu\sigma\rho} \,\partial_\mu A_{\nu\sigma\rho} \Big)
 - \,\dfrac{1}{4} \, \big( B_{\mu\nu}^{(1)} \big)^2 + \dfrac{1}{2}\, B_{\mu\nu}^{(1)}\, \big(\partial_\sigma A^{\sigma\mu\nu}\big)   \nonumber\\
&-& \dfrac{1}{4}\, \big( B_{\mu\nu}^{(2)} \big)^2 + \dfrac{1}{2} \, B_{\mu\nu}^{(2)}\, \Big( \varepsilon^{\mu\nu\sigma\rho} \,\partial_\sigma A_\rho \Big)  
  - B \, (\partial \cdot A) + \dfrac{1}{2}\, B^2,
\end{eqnarray}
respects the discrete duality symmetry transformations: $A_\mu \rightarrow \mp \,(1/3!)\, \varepsilon_{\mu\nu\sigma\rho}\, A^{\nu\sigma\rho},\, A_{\mu\nu\sigma} \rightarrow \pm\, \varepsilon_{\mu\nu\sigma\rho}\, A^\rho,\, B \rightarrow \mp\, B_1, \, B_1 \rightarrow 
\pm\, B, \, B_{\mu\nu}^{(1)} \rightarrow \pm\, B_{\mu\nu}^{(2)}, \, B_{\mu\nu}^{(2)} \rightarrow \mp\,  B_{\mu\nu}^{(1)}$.
Second, the linearized Lagrangian density (7) will be {\it further} generalized (i.e. $ {\cal L}_{(2)} \to {\cal L}_{(3)}$)
by incorporating a polar-vector field ($\phi_\mu $) and an axial-vector field ($\tilde \phi_\mu $) in the next section.\\

\vskip 0.7cm

\section{Nilpotent (co-)BRST Symmetry Transformations}

A more general and linearized form of the Lagrangian density for the free Abelian 3-form gauge theory has been worked out in our earlier work [19].
This  Lagrangian density ${\cal L}_{(3)} $ incorporates the (axial-)vector fields $(\tilde \phi_\mu)\phi_\mu$  at appropriate places as follows
\begin{eqnarray}\label{8}
{\cal L}_{(3)}  &=& \dfrac{1}{2}\, B^2 - B \, (\partial \cdot A) +  \dfrac{1}{2} \, B_1^2 -\, B_1\, \Big(\dfrac{1}{3!}\varepsilon^{\mu\nu\sigma\rho} \,\partial_\mu A_{\nu\sigma\rho} \Big) \nonumber\\
 &-& \,\dfrac{1}{4} \, \big( B_{\mu\nu}^{(1)} \big)^2 + \dfrac{1}{2}\, B_{\mu\nu}^{(1)}\, \Big[\partial_\sigma A^{\sigma\mu\nu}
+ \dfrac{1}{2}\, \big (\partial^\mu \phi^\nu - \partial^\nu \phi^\mu \big) \Big ] - \dfrac{1}{4}\, B_2^2 + 
\dfrac{1}{2}\, B_2 \big (\partial \cdot \phi \big )   \nonumber\\
&-& \dfrac{1}{4}\, \big( B_{\mu\nu}^{(2)} \big)^2 + \dfrac{1}{2} \, B_{\mu\nu}^{(2)}\, \Big[ \varepsilon^{\mu\nu\sigma\rho} \,\partial_\sigma A_\rho 
+ \dfrac{1}{2}\, \big (\partial^\mu \tilde \phi^\nu - \partial^\nu \tilde \phi^\mu \big) \Big] - \dfrac{1}{4}\, B_3^2 + 
\dfrac{1}{2}\, B_3 \big (\partial \cdot \tilde \phi \big ), 
\end{eqnarray}
where  the additional set of bosonic Nakanishi-Lautrup type auxiliary fields ($B_2, \, B_3 $) have been invoked to linearize the gauge-fixing terms for the 
{\it additional} polar-vector ($\phi_\mu$) and axial-vector  ($\tilde \phi_\mu$) fields.
It is straightforward to check that
the above linearized version of the Lagrangian density ${\cal L}_{(3)}$ respects the following set of discrete duality transformations
\begin{eqnarray}\label{9}
&&A_\mu \longrightarrow \,\mp\, \dfrac{1}{3!}\, \varepsilon_{\mu\nu\sigma\rho}\, A^{\nu\sigma\rho}, \quad
A_{\mu\nu\sigma} \longrightarrow \,\pm\, \varepsilon_{\mu\nu\sigma\rho}\, A^\rho, \quad B_{\mu\nu}^{(1)} \to \pm\, B_{\mu\nu}^{(2)}, \quad
B_{\mu\nu}^{(2)} \to \mp\, B_{\mu\nu}^{(1)}, \nonumber\\
&& B \to \mp\, B_1, \quad B_1 \to \pm\, B, \quad B_2 \to \pm\,  B_3, \quad B_3 \to \mp\, B_2, \;\; \phi_\mu \to \pm\,  \tilde \phi_\mu, \; \;
\tilde \phi_\mu \to \mp\, \phi_\mu,
\end{eqnarray}
which is the generalization of such transformations that have been
mentioned after equation (7). The Faddeev-Popov (FP) ghost terms 
for the free Abelian 3-form gauge theory have been obtained
in our earlier work [19] and we have the {\it standard} FP-ghost term for the Abelian 1-form theory. The full form of the FP-ghost part of the 
Lagrangian density ${\cal L}_{(FP)} $, in addition to the properly gauge-fixed Lagrangian density ${\cal L}_{(3)} $, for 
our BRST-quantized combined 4D field-theoretic system of the Abelian 3-form and 1-form gauge theory\footnote{Besides a few
changes in the notations, we have taken an overall factor of {\it half}  outside the square bracket
of the FP-ghost terms that have been taken in our earlier work on the BRST approach to the description of the free Abelian 3-form theory [19] because
we note that this difference of overall factor is present in our gauge-fixed Lagrangian density (2) which respects the 
discrete duality symmetry transformations (3). The Lagrangian density ${\cal L}_B = {\cal L}_{(3)} + {\cal L}_{(FP)} $
describes the {\it combined} 4D field-theoretic system of the free
Abelian 3-form and 1-form {\it massless} gauge theories within the framework of BRST formalism.} is [19]\\
\begin{eqnarray}\label{10}
{\cal L}_{(FP)}  &=&  \dfrac{1}{2}\, \Big [\big (\partial_\mu \bar C_{\nu\sigma} +  \partial_\nu \bar C_{\sigma\mu} 
+ \partial_\sigma \bar C_{\mu\nu} \big ) \big (\partial^\mu C^{\nu\sigma} \big ) +
\big (\partial_\mu \bar C^{\mu\nu}  + \partial^\nu \bar C_1 \big ) f_\nu \nonumber\\ 
&-& \big (\partial_\mu  C^{\mu\nu}  + \partial^\nu  C_1 \big ) \bar F_\nu 
 +  \big( \partial \cdot \bar \beta \big) \, B_4
 - \big( \partial \cdot  \beta \big) \, B_5 - B_4 \, B_5 - 2\, \bar F^\mu\, f_\mu \nonumber\\
&-& \big (\partial_\mu \bar \beta_\nu - \partial_\mu \bar \beta_\nu \big ) 
\big (\partial^\mu \beta^\nu \big )  
-\, \partial_\mu \bar C_2\, \partial^\mu C_2  \Big ] - \partial_\mu \bar C\, \partial^\mu C, 
\end{eqnarray}
where the fermionic (anti-)ghost fields $(\bar C)C$, present in the last term, 
are associated with the Abelian 1-form gauge field $A_\mu$ and they carry the ghost numbers
(-1)+1, respectively. On the other hand, corresponding to our Abelian 3-form gauge field $A_{\mu\nu\sigma} $, we have the antisymmetric
($\bar C_{\mu\nu} = -\, \bar C_{\nu\mu}, C_{\mu\nu} = -\,  C_{\nu\mu}$) tensor (anti-)ghost fields $(\bar C_{\mu\nu}) C_{\mu\nu}$ which are endowed 
with the ghost numbers (-1)+1, respectively. In our theory, we have 
the ghost-for-ghost {\it bosonic} vector (anti-)ghost fields $(\bar \beta_\mu)\beta_\mu$
and the ghost-for-ghost-for-ghost {\it fermionic} (anti-)ghost fields $(\bar C_2)C_2$ that carry the ghost numbers (-2)+2 and (-3)+3., respectively. 
The fermionic auxiliary fields $(\bar F_\mu)f_\mu$ and bosonic auxiliary fields $(B_5)B_4$ of our theory carry the ghost numbers
(-1)+1 and (-2)+2, respectively. The additional (anti-)ghost fields $(\bar C_1)C_1$ are endowed with the ghost numbers (-1)+1, respectively.
The above FP-ghost part of the Lagrangian density (10) respects the following discrete symmetry transformations:
\begin{eqnarray}\label{11}
&& C_{\mu\nu} \longrightarrow \,\pm\,\bar C_{\mu\nu}, \quad \bar C_{\mu\nu} \longrightarrow \,\mp\, C_{\mu\nu}, \quad
\beta_\mu \to \pm\,  \bar \beta_\mu, \quad \bar \beta_\mu \to \mp\,  \beta_\mu, \quad 
f_\mu \to \pm\, \bar F_\mu, \nonumber\\
&& \bar F_\mu \to \mp\,  f_\mu, \qquad B_4 \to \mp\,  B_5, \qquad B_5 \to \pm\,  B_4,
\qquad C \to  \mp\, \bar C, \qquad \bar C \to \pm\,   C, \nonumber\\
&&  C_2 \to \pm\,  \bar C_2, \qquad 
\bar C_2 \to \mp\,   C_2, \qquad C_1 \to \pm\,  \bar C_1, \qquad 
\bar C_1 \to \mp\,   C_1.
\end{eqnarray}
Thus, we note that the total Lagrangian density ${\cal L}_{(B)} = {\cal L}_{(3)} + {\cal L}_{(FP)}$ 
[which is the sum of (8) and (10)] remains invariant under the 
discrete symmetry transformations (9) and (11).

We focus now on a few useful continuous symmetry transformations of the total Lagrangian density ${\cal L}_{(B)}$. In this connection, it is interesting
to point out that the following infinitesimal and off-shell nilpotent (i.e. $s_{(d)b}^2 = 0 $) (co-)BRST transformations ($s_{(d)b}$)
\begin{eqnarray}\label{12}
&& s_d A_{\mu \nu\sigma} = \varepsilon_{\mu\nu\sigma\rho}\, \partial^\rho \bar C, 
\quad 
s_d A_\mu = \dfrac{1}{2} \,\varepsilon_{\mu\nu\sigma\rho}\,\partial^\nu \bar C^{\sigma\rho}, \quad s_d  \bar C_{\mu\nu}  =  \partial_\mu \bar \beta_\nu - \partial_\nu \bar \beta_\mu, \nonumber\\
&& s_d \bar \beta_\mu = \partial_\mu \bar C_2, \qquad\;\;  s_d  C_1 = -\,B_3 , \qquad \;\;s_d \beta_\mu = -\, f_\mu,  
\qquad\;\; s_d \tilde \phi_\mu  = +\, \bar F_\mu, \nonumber\\
&&  s_d  C_{\mu\nu} = - \, B_{\mu\nu}^{(2)},    \qquad s_d C = -\, B_1, \qquad
s_d C_2 = B_4, \qquad s_d \bar C_1 = B_5, \nonumber\\
&&s_d \; \Bigl [ \bar C_2,\, \bar C,\, f_\mu,\, {\bar F}_\mu,\, \phi_\mu, \, B,\, B_1, \, B_2,\, B_3, \,  B_4,\, B_5,\, B^{(1)}_{\mu\nu}, \, B^{(2)}_{\mu\nu}
\Bigl ]\; = \;0, 
\end{eqnarray}
\begin{eqnarray}\label{13}
&&s_b A_{\mu\nu\sigma} = \partial_\mu C_{\nu\sigma} + \partial_\nu C_{\sigma\mu}
+ \partial_\sigma C_{\mu\nu}, \quad  s_b C_{\mu\nu} = \partial_\mu \beta_\nu
- \partial_\nu \beta_\mu, \quad s_b \bar C_{\mu\nu} = B^{(1)}_{\mu\nu}, \nonumber\\
&&s_b A_{\mu} = \partial_\mu C, \qquad s_b \bar C = B, \quad  
s_b \bar \beta_\mu = \bar F_\mu, \qquad
s_b \beta_\mu = \partial_\mu C_2,  \nonumber\\
&&s_b {\bar C}_2 = B_5, \qquad s_b C_1 = - B_4, \qquad  s_b \bar C_1 = B_2, \qquad s_b \phi_\mu = f_\mu, 
\nonumber\\
&&s_b \; \Bigl [ C_2,\, C,\, f_\mu,\, {\bar F}_\mu,\,\tilde \phi_\mu, \, B,\, B_1, \, B_2,\, B_3, \,  B_4,\, B_5,\, B^{(1)}_{\mu\nu}, \, B^{(2)}_{\mu\nu}
\Bigl ]\; = \;0,
\end{eqnarray}
leave the action integral, corresponding to the Lagrangian density ${\cal L}_{(B)}$, invariant because we observe that
{\it this} Lagrangian density transforms to the total spacetime derivatives as: 
\begin{eqnarray}\label{14}
s_d {\cal L}_{(B)} &=& \dfrac{1}{2}\, \partial_\mu\, \Big[ (\partial^\mu\, \bar C^{\nu\sigma} + \partial^\nu\, \bar C^{\sigma\mu}
+ \partial^\sigma\, \bar C^{\mu\nu}) \, B^{(2)}_{\nu\sigma}  + B^{\mu\nu(2)}\, \bar F_\nu + B_4\, \partial^\mu\,\bar C_2  \nonumber\\
&+& B_5\, f^\mu + B_3\, \bar F^\mu + (\partial^\mu\, \bar \beta^\nu - \partial^\nu\, \bar \beta^\mu)\, f_\nu \Big] 
- \partial_\mu \Big [B_1 \, \partial^\mu \bar C \Big ],
\end{eqnarray}
\begin{eqnarray}\label{15}
s_b\, {\cal L}_{(B)} &=& \dfrac{1}{2}\, \partial_\mu\, \Big[ (\partial^\mu\, C^{\nu\sigma} + \partial^\nu\, C^{\sigma\mu}
+ \partial^\sigma\, C^{\mu\nu}) \, B^{(1)}_{\nu\sigma}  + B^{\mu\nu(1)}\, f_\nu - B_5\, \partial^\mu\, C_2  \nonumber\\
&+& B_2\, f^\mu + B_4\, \bar F^\mu - (\partial^\mu\,  \beta^\nu - \partial^\nu\, \beta^\mu)\,  \bar F_\nu \Big] 
- \partial_\mu \Big [B \, \partial^\mu  C \Big ].
\end{eqnarray}
Thus, we draw the conclusion that the infinitesimal and off-shell nilpotent (co-)BRST transformations [cf. Eqs. (12),(13)] are the {\it symmetry}
transformations for our present combined 4D field-theoretic system  of the free Abelian 3-form and 1-form gauge theories.

We end this section with a few remarks. We note that the total kinetic terms of {\it all} the basic fields  
(owing their origin to {\it primarily} exterior derivative of differential geometry) remain invariant under
the  off-shell nilpotent BRST symmetry transformations. On the other hand, under the  nilpotent co-BRST symmetry transformations, the total gauge-fixing terms
for {\it all} the basic fields
(tracing their existence {\it basically} to the co-exterior derivative of differential geometry) 
 remain unchanged\footnote{In a recent paper [22], we find the discussion on the 
nilpotent co-BRST symmetry transformations where the gauge-fixing term remains invariant. However, we do {\it not} find  any logical
and/or mathematical 
explanation for such an invariance.}. There are physical consequences of the {\it newly} introduced set of
(axial-)vector fields (cf. Conclusions for more details).\\

\vskip 0.9cm

\section{ Bosonic Symmetry and Algebraic Structures of the Continuous and Discrete Symmetry Operators}

The anticommutator (i.e. $\{ s_b, \; s_d \} $) between the off-shell nilpotent 
versions of symmetries in our 
equations (12)  and (13) is {\it not} equal to zero. In fact, this anticommutator defines a set of
a non-nilpotent bosonic symmetry (i.e. $s_\omega = \{ s_b, \; s_d \} $) transformations ($s_\omega$), under which, the
Lagrangian density ${\cal L}_{(B)} $ transforms to the total spacetime derivative thereby rendering the 
action integral (corresponding to {\it this} Lagrangian density) invariant. To corroborate this statement, 
we take recourse to (i) our observations in (14) and (15), (ii) use the infinitesimal and off-shell  nilpotent (co-)BRST symmetry
transformations ($s_{(d)b}$) of equations (12) as well as (13), and 
exploit the straightforward {\it definition} of the infinitesimal bosonic symmetry transformation
operator: $s_\omega = \{ s_b, \; s_d \} \equiv s_b \, s_d + s_d\, s_b$. Mathematically, this whole operation can be 
succinctly expressed as:
\begin{eqnarray}\label{16}
&&s_\omega \, {\cal L}_{(B)} =  \big ( s_b \, s_d + s_d \, s_b \big )\, {\cal L}_{(B)} \nonumber\\
&&\equiv \dfrac{1}{2}\, \partial_\mu\, \Big[ \big \{\partial^\mu\, B^{\nu\sigma(1)} + \partial^\nu\, B^{\sigma\mu(1)}
+ \partial^\sigma\, B^{\mu\nu(1)} \big \} \, B^{(2)}_{\nu\sigma}  \nonumber\\
&& - \,\big \{ \partial^\mu\, B^{\nu\sigma(2)} + \partial^\nu\, B^{\sigma\mu(2)}
+ \partial^\sigma\, B^{\mu\nu(2)} \big \} \, B^{(1)}_{\nu\sigma} + B_4 \, \partial^\mu B_5 - B_5 \, \partial^\mu B_4 \nonumber\\
&& + \,\big (\partial^\mu f^\nu - \partial^\nu f^\mu \big )\, \bar F_\nu - \big (\partial^\mu \bar F^\nu - \partial^\nu \bar F^\mu \big )\, f_\nu
\Big ] + \partial_\mu \Big [\big (B\, \partial^\mu B_1 - B_1\, \partial^\mu B \big ) \Big ].
\end{eqnarray}
The above transformation of the Lagrangian density ${\cal L}_{(B)} $ can {\it also} be obtained from the operation of the non-nilpotent bosonic 
symmetry operator $s_\omega$ on the 
{\it individual} field of this Lagrangian density. In other words, the following field transformations
under $s_\omega$, namely;
\begin{eqnarray}\label{17}
&& s_\omega A_{\mu\nu\sigma} =  \varepsilon_{\mu\nu\sigma\rho}\, \partial^\rho B  -
\Big (\partial_\mu\, B^{(2)}_{\nu\sigma} + \partial_\nu\, B^{(2)}_{\sigma\mu}
+ \partial_\sigma\, B_{\mu\nu}^{(2)} \Big ),   \nonumber\\
&& s_\omega A_\mu = \dfrac{1}{2} \, \varepsilon_{\mu\nu\sigma\rho}\, \partial^\nu B^{\sigma\rho(1)} - \partial_\mu B_1, 
\qquad s_\omega \bar \beta_\mu = \partial_\mu B_5, \qquad s_\omega \beta_\mu = \partial_\mu  B_4, \nonumber\\
&& s_\omega C_{\mu\nu} = -\, \big (\partial_\mu f_\nu - \partial_\nu f_\mu \big ), \qquad 
s_\omega \bar C_{\mu\nu} = +\, \big (\partial_\mu \bar F_\nu - \partial_\nu \bar F_\mu \big ), \nonumber\\
&& s_\omega \Big[B, \, B_1,  B_2,  B_3,  B_4,  B_5, \, \phi_\mu,\, \tilde \phi_\mu,\, f_\mu, \, \bar F_\mu,\, 
C, \bar C, \, C_1,  \bar C_1,  C_2,  \bar C_2, B_{\mu\nu}^{(1)}, \, B_{\mu\nu}^{(1)}  \Big] = 0,
 \end{eqnarray}
{\it also} lead to the derivation of (16). 
At this stage, it is worthwhile to mention that under the above bosonic symmetry transformations, the (anti-)ghost fields {\it either} do not 
transform at all {\it or} they transform up to the $U(1)$ gauge symmetry-type transformations.

It is interesting to point
out that, in their operator forms, the  (co-)BRST transformations $s_{(d)b} $ and the bosonic transformation $s_\omega $ obey the
following algebra, namely;
\begin{eqnarray}\label{18}
 &&s_b^2 = 0, \quad \qquad s_d^2 = 0, \quad \qquad s_\omega = \big \{ s_b, \; s_d \big \} \equiv \big (s_b + s_d \big )^2, \nonumber\\
 && \big [s_\omega, \; s_b \big ] = 0, \quad \qquad 
\big [s_\omega, \; s_d \big ] = 0, \quad \qquad  \big \{ s_b, \; s_d \big \} \neq 0,
\end{eqnarray} 
which establish that the non-nilpotent bosonic symmetry transformation, in its operator form, commutes with {\it both} the 
off-shell nilpotent (co-)BRST symmetry transformation operators. This  can be proved in a very simple manner by taking into account the
off-shell nilpotency ($s_{(d)b}^2 = 0 $)
of the (co-)BRST symmetry transformation operators $s_{(d)b} $ and the straightforward definition 
(i.e. $s_\omega = s_b\, s_d + s_d\, s_b $) of the non-nilpotent bosonic symmetry
transformation operator $s_\omega$.
The algebra (18) resembles with  the 
following algebra obeyed by a set of {\it three}  de Rham cohomological operators of differential geometry [15,16]
\begin{eqnarray}\label{19}
&&d^2 = 0, \quad \qquad \delta^2 = 0, \quad \qquad \Delta = \big \{ d, \; \delta \big \} \equiv \big (d + \delta\big )^2 , \nonumber\\
&& \big [\Delta, \; d \big ] = 0, \quad \qquad \big [\Delta, \; \delta \big ] = 0, \quad \qquad  \big \{ d, \; \delta \big \} \neq 0,
\end{eqnarray}
where $d$ (with $d^2 = 0$) is the exterior derivative, $\delta = \pm\, *\, d\, *$ (with $\delta^2 = 0$) is the 
co-exterior (or dual-exterior) derivative and $\Delta = (d + \delta)^2$ is the Laplacian operator. 
Here the mathematical symbol $*$ denotes the Hodge duality operator on a given  spacetime 
manifold on which the cohomological operators are defined (see, e.g. [15,16] for details).

The uncanny resemblance between the algebraic structures (18) and (19) establishes that we have obtained the physical realization of the
abstract mathematical objects (like the cohomological operators of differential geometry [15,16] because we have the mapping: $s_b \Leftrightarrow d, \;
s_d \Leftrightarrow \delta,\; s_\omega \Leftrightarrow \Delta $). However, we have {\it not} discussed the anti-BRST, anti-co-BRST and ghost-scale
symmetries in our present investigation. Hence, the above mapping is {\it not} complete  yet. We have obtained the one-to-one mapping 
because we have considered {\it only} the Lagrangian density ${\cal L}_{(B)} = {\cal L}_{(3)} + {\cal L}_{(FP)} $ [cf. Eqs. (8),(10)]
at the {\it quantum} level which respects the kinds of symmetries that we have focused in our present endeavor. There exists a
 possibility of having a {\it coupled} (but equivalent) version of the quantum Lagrangian density that respects the 
anti-BRST and anti-co-BRST symmetries. If we had  considered the other {\it quantum} version of the coupled Lagrangian density along with ${\cal L}_{(B)}$,  
we would have ended up with the two-to-one 
mapping between the symmetry transformation operators and the cohomological operators as we have obtained 
in our earlier works (see, e.g.  [8,9,17]).

Physically, the above one-to-one mapping (i.e. $s_b \Leftrightarrow d, \;
s_d \Leftrightarrow \delta,\; s_\omega \Leftrightarrow \Delta $) is meanigful because we observe that the kinetic terms of the basic fields
(owing their origin to the exterior derivative $d$) remain invariant under the 
nilpotent BRST transformation operator $s_b$. On the other hand, the
gauge-fixing terms (originating from the operation of the co-exterior derivative $\delta$ on the basic fields) remain unchanged under the 
nilpotent co-BRST
transformations  $s_d$. As far as the non-nilpotent bosonic symmetry transformation operator $s_\omega$ is concerned, we note that (i) the
(anti-)ghost fields of our theory {\it either} do not transform at all {\it or} transform up to a U(1) gauge symmetry-type transformation under it, and
(ii) it commutes with the off-shell nilpotent (anti-)co-BRST symmetry operators. We have not yet provided
the physical realization of the 4D algebraic relationship: $\delta = - \, *\, d \, * $ that exists between the (co-)exterior derivatives $(\delta)d $
of differential geometry [15,16]. In the next paragraph, we accomplish this goal in terms of the interplay between the discrete and continuous symmetry
transformation operators of our 4D field-theocratic system.

Against the backdrop of the above paragraph, first of all, we note that the mathematical relationship:  $\delta = - \, *\, d \, * $ is true for
any {\it even} dimensional spacetime manifold (including 4D) where, as is well-known, the (co-)exterior derivatives $(\delta)d$ are nilpotent
(i.e. $\delta^2 = 0, \, d^2 = 0 $) of order two. In the context of our present 4D BRST-quantized field-theocratic model, interestingly, we have two
off-shell nilpotent (i.e. $s_{(d)b}^2 = 0 $) continuous (co-)BRST symmetry transformation operators $s_{(d)b}$. On the other hand, we also have a set of 
discrete duality symmetry transformations in (9) and (11) in the (non-)ghost sectors of the Lagrangian density 
${\cal L}_{(B)} = {\cal L}_{(3)} + {\cal L}_{(FP)} $ in our theory, too. We find that the interplay between the continuous and discrete symmetry 
transformation operators provide the physical realization of the mathematical relationship: $\delta = - \, *\, d \, * $ in the following manner
\begin{eqnarray}\label{20}
 s_d \, \Phi = -\, *\, s_b \, *\, \Phi, \qquad \; \Phi &=& A_{\mu\nu\sigma}, B_{\mu\nu}^{(1)}, B_{\mu\nu}^{(2)}, \bar C_{\mu\nu},  C_{\mu\nu}, A_\mu, 
\phi_\mu,  \tilde \phi_\mu, f_\mu, \bar F_\mu, 
\bar \beta_\mu,  \beta_\mu, \nonumber\\
&& \bar C,  C,  \bar C_1,  C_1, \bar C_2,  C_2,  B, B_1, B_2, B_3, B_4, B_5, 
\end{eqnarray}
where the symbol $*$ stands for the discrete duality symmetry transformations.  In the above equation (20), as is obvious, the generic 
field of the Lagrangian density ${\cal L}_{(B)} $ has been denoted by the field $\Phi$. The ($-$) sign, 
on the r.h.s. of the above equation (20), is dictated by a couple of successive operations 
of the discrete duality symmetry transformation operators [cf. Eqs. (9),(11)]  on the generic field $\Phi$ of the Lagrangian 
density ${\cal L}_{(B)} $ as [23]:
\begin{eqnarray}\label{21}
 * \; \big (*\; \Phi \big ) = -\, \Phi.
 \end{eqnarray}
Let us take a couple of fields from the (non-)ghost sectors of the Lagrangian density ${\cal L}_{(B)} $ to corroborate our above claims.
First of all, from equation (12), it is clear that $s_d A_\mu = \frac{1}{2} \varepsilon_{\mu\nu\sigma\rho}\,\partial^\nu \bar C^{\sigma\rho} $. On the
other hand, the relationship (20) implies that we have: $s_d A_\mu = -\, *\, s_b\, * A_\mu$. In what follows, we carry out the explicit
evaluation of the r.h.s (i.e. $-\, *\, s_b\, * A_\mu $) of this relationship for the sake of readers' convenience. The explicit computation is:
$
 -\, *\, s_b\, * A_\mu = \pm\, \frac{1}{3!}\, \varepsilon_{\mu\nu\sigma\rho}\, *\, s_b A^{\nu\sigma\rho} \equiv 
 \pm\, \frac{1}{3!}\, \varepsilon_{\mu\nu\sigma\rho}\, *\, \big (\partial^\nu C^{\sigma\rho} 
+ \partial^\sigma C^{\rho\nu} + \partial^\rho C^{\nu\sigma} \big ) 
 \equiv \frac{1}{3!}\, \varepsilon_{\mu\nu\sigma\rho}\,  
\big (\partial^\nu \bar C^{\sigma\rho} 
+ \partial^\sigma \bar C^{\rho\nu} + \partial^\rho \bar C^{\nu\sigma} \big ) =  \frac{1}{2}\, \varepsilon_{\mu\nu\sigma\rho}\,\partial^\nu \bar C^{\sigma\rho},
$
where we have used (i) the discrete duality symmetry transformations from (9) and (11), and (ii) the appropriate BRST symmetry
transformation from (13). In exactly similar fashion, it is straightforward to verify that $s_d C_{\mu\nu} = -\, B_{\mu\nu}^{(2)}$ can be derived
from: $ -\,*\, s_b\, *\, C_{\mu\nu}$ by taking into account the discrete  duality symmetry transformations from (9) and (11) and the appropriate
continuous BRST symmetry transformation from (13). In other words, we have: $ -\,*\, s_b\, *\, C_{\mu\nu} 
= \mp\, * s_b \bar C_{\mu\nu} = \mp\, * B_{\mu\nu}^{(1)} = -\, B_{\mu\nu}^{(2)}$.
Thus, we conclude that the Hodge duality $*$ operator can be physically realized in terms of 
the discrete duality symmetry transformations [cf. Eqs. (9),(11)] that
are present  in the (non-)ghost sectors of our Lagrangian  density  ${\cal L}_{(B)} $. On the other hand, the 
nilpotent (i.e. $\delta^2 = 0, \; d^2 = 0 $) (co-)exterior derivatives $(\delta)d $
can be given their physical meaning in terms of the off-shell nilpotent ($s_{(d)b}^2 = 0 $)
(co-)BRST symmetry transformation operators $s_{(d)b}$. Thus, we have been able to provide the physical realization of the 
mathematical relationship: $\delta = -\, *\, d\, * $ between the (co-) exterior derivatives $(\delta)d$ in terms of the interplay
between the discrete and continuous symmetry operators of our present 4D field-theoretic example for Hodge theory.\\

\vskip 0.7cm

\section{Conclusions}

In our present investigation, we have provided the physical realization of the {\it abstract} algebraic structures that are obeyed 
by the well-known de Rham cohomological operators of differential geometry [15,16] in the terminology of the  {\it two} off-shell nilpotent BRST and
co-BRST (i.e. dual-BRST) symmetry transformation operators and a non-nilpotent bosonic symmetry transformation operator that is derived from 
the anticmmutator of the {\it above} off-shell nilpotent (co-)BRST symmetry transformation operators. It is worthwhile to point out that
the bosonic symmetry transformation operator commutes with {\it both} the nilpotent 
BRST and dual-BRST (i.e. co-BRST) symmetry transformation operators of
our present 4D BRST-quantized field-theoretic model of the Abelian 3-form and 1-form gauge theories. This observation is exactly like
the algebraic structure (19) where the celebrated Laplacian operator $\Delta$ 
of the cohomological operators commutes with the nilpotent ($d^2 = \delta^2 = 0 $)
(co-)exterior derivatives $(\delta)d$ of differential geometry [15,16].

We have laid a great deal of emphasis on the existence of the discrete duality symmetry transfigurations [cf. Eq. (9)] in the non-ghost sector and 
discrete symmetry transformations [cf. Eq. (11)] in the ghost sector of the Lagrangian density ${\cal L}_{(B)}$ (cf. the second and third sections) because
{\it these} symmetry transformation operators provide the physical realization of the Hodge duality $*$ operator of differential geometry in the
mathematical relationship: $\delta = - \, *\, d\, * $ between the (co-)exterior [i.e. (dual-)exterior] derivatives. The relationship
between the Abelian 1-form and 3-form basic gauge fields in (9) 
establishes that there is an explicit duality between these two {\it basic} gauge fields when they are present {\it together}
in a 4D field-theoretic model of Hodge theory. This is one of the highlights of our present endeavor (where the {\it basic} gauge fields
of two {\it different} Abelian gauge theories are related to each-other by a set of discrete duality symmetry transformations when 
{\it these} theories are taken {\it together} in a single combined 4D field-theoretic system).

As far as the physical consequences of our present investigation are concerned, we would like to pinpoint our observation that there is appearance of the
vector (i.e. $\phi_\mu$) and axial-vector (i.e. $\tilde \phi_\mu$) fields in our theory on the symmetry grounds {\it alone}. It turns out that both these
basic fields appear with the {\it positive} kinetic terms which is a {\it unique} feature of our present field-theoretic example for Hodge theory.
Unlike our present system, we have been able to establish (see, e.g. [8,9,19] and references therein)
that the Abelian $p$-form (i.e. $p = 1, 2, 3 $) massless and 
St{\" u}ckelberg-modified massive gauge theories in the D $= 2 p$ (i.e. D = 2, 4, 6) dimensions of spacetime are 
the tractable field-theoretic examples for Hodge theory where there is {\it always} appearance of the 
``exotic'' fields with the {\it negative} kinetic terms. In a very recent work [24], we have been able to show the existence of a 
massless pseudo-scalar field (with the negative kinetic term) in an odd dimensional (i.e. 3D) field-theoretic example for Hodge theory.
One of the highlights of our present endeavor is the observation that
such kinds of ``exotic'' fields do {\it not} appear in our present BRST-quantized 4D field-theoretic example for Hodge theory. This result is indeed 
a {\it novel} observation
 in our present investigation vis-{\` a}-vis our earlier works on the field-theoretic models of Hodge theory 
within the framework of BRST formalism (see, e.g. [8,9,17-19] for details).

At this juncture, we would like to compare and contrast (in an elaborate manner) our observations in the context of the 4D BRST-quantized field-theoretic 
examples for Hodge theory in our earlier work [9] and present work. The {\it former} 
BRST-quantized 4D theory  is the St{\" u}ckelberg-modified massive Abelian 2-form theory [9]. On the other hand, our present 4D field-theoretic system is a
combination of the free Abelian 3-form and 1-form gauge theories within the framework of BRST formalism. We would like to lay emphasis on
the fact that the axial-vector field appears in {\it both} the BRST-quantized 4D  theories. However, there is a discerning
difference as far as the kinetic terms associated with the axial-vector field are concerned\footnote{Both the  4D field-theoretic examples for Hodge theory
incorporate the polar-vector field, too. However, {\it this} field turns up with the positive kinetic term in {\it both} the above theories. The PS field 
does {\it not} appear in our present field-theoretic system. Hence, we do {\it not} comment anything on this field.}. 
In our earlier work [9], the axial-vector is endowed
with a {\it negative} kinetic term but it carries a {\it positive} kinetic term in our present endeavor. Thus, the axial-vector field {\it cannot}
be  a {\it true} candidate for the phantom field of cosmology and one of the possible candidates for dark energy/dark matter. On the other hand,
we have seen that the pseudo-scalar (PS) field  carries the {\it negative} kinetic term  in the 
BRST-quantized field-theoretic models of (i) the 2D {\it modified}
Proca (i.e. a massive Abelian 1-form) theory [17], (ii) the 3D combined system of the Abelian 2-form and 1-form gauge theories [18], and
(iii) the 4D {\it modified} massive Abelian 2-form theory [9]. Hence, as far as our present and earlier works [8,9,17,18,19]  are concerned,
we are convinced that that the PS field is the most fundamental field which (i) corresponds to the phantom field of cosmology,
and (ii) represents one of the possible candidates for the dark matter/dark energy.

In our future endeavor, we wish to discuss the coupled (but equivalent) Lagrangian densities, nilpotent (anti-)BRST and (anti-)co-BRST symmetries,
a unique bosonic symmetry and the ghost-scale symmetry {\it along} with the a couple of useful discrete symmetries to obtain the physical realization(s) of the
cohomological operators in terms (i) the symmetry operators, and (ii) the {\it appropriate} conserved charges, at the algebraic level. We have
{\it not} discussed anything about the Curci-Ferrari (CF) type restrictions which are the hallmark of a properly BRST-quantized theory.  We are
sure that our present theory would be endowed with the {\it non-trivial} CF-type restrictions (see, e.g. [25] for details). 
As a consequence, the {\it Noether} (anti-)BRST
charges would turn out to be non-nilpotent. We plan to derive the off-shell nilpotent versions of the conserved (anti-)BRST charges
and discuss the physicality criteria w.r.t. to the off-shell nilpotent versions of {\it these} charges 
(following the theoretical technique proposed in our earlier work [24]) and demonstrate that
the physical states (exiting in the {\it total} quantum Hilbert space of states) are annihilated by the operator forms of the 
first-class constraints of our present  4D {\it classical} gauge theories. We shall also derive the BRST algebra with the off-shell nilpotent versions of the
(anti-)BRST and (anti-) co-BRST charges and the {\it other} conserved charges of our theory (which would be shown to be the generators for the 
{\it six} continuous symmetry transformations of our BRST-quantized 4D field-theoretic example for the Hodge theory). \\

\noindent
{\bf Data Availability Statement}\\

\noindent
No new data were created or analyzed in this study.\\

\noindent
{\bf Conflicts of interest}\\ 

\noindent
The author declares that there are no conflicts of interest.\\

\noindent
{\bf Funding Statement}\\

\noindent
No funding was received for this research.\\

\end{document}